\documentclass[conference]{IEEEtran}
\IEEEoverridecommandlockouts
\usepackage{amsmath,amssymb,amsfonts}
\usepackage{algorithmic}
\usepackage{graphicx}
\usepackage{textcomp}
\usepackage{xcolor}
\usepackage{bm}
\usepackage{multirow}
\usepackage{booktabs}
\usepackage{makecell}
\usepackage[ruled,linesnumbered]{algorithm2e}
\usepackage{setspace}
\usepackage{graphicx}
\graphicspath{{figs/}}
\usepackage{subfigure} 
\usepackage{amsmath}
\usepackage{amsthm}
\usepackage{float}
\usepackage{hyperref} 
\hypersetup{hidelinks} 
\usepackage{tikz}
\usepackage[numbers,sort&compress]{natbib}
\usepackage{balance}
\usetikzlibrary{tikzmark,calc}
\usepackage{placeins} 
\def\BibTeX{{\rm B\kern-.05em{\sc i\kern-.025em b}\kern-.08em
		T\kern-.1667em\lower.7ex\hbox{E}\kern-.125emX}}
\newcommand{\mytitle}{Cali3F: Calibrated Fast Fair Federated Recommendation System}
\DeclareRobustCommand*{\IEEEauthorrefmark}[1]{
	\raisebox{0pt}[0pt][0pt]{\textsuperscript{\footnotesize\ensuremath{#1}}}}

\begin{document}
	\title{\mytitle}
	\author{\IEEEauthorblockN{Zhitao Zhu\IEEEauthorrefmark{1,2}, Shijing Si\IEEEauthorrefmark{1}, Jianzong Wang\IEEEauthorrefmark{1*}, Jing Xiao\IEEEauthorrefmark{1}}
		\IEEEauthorblockA{\IEEEauthorrefmark{1}\textit{Ping An Technology (Shenzhen) Co., Ltd.}, Shenzhen, China\\
			\IEEEauthorblockA{\IEEEauthorrefmark{2}\textit{Institude of Advanced Technology, University of Science and Technology of China}, Hefei, China}		
			Emails: andyzzt@mail.ustc.edu.cn, shijing.si@outlook.com, jzwang@188.com, xiaojing661@pingan.com.cn}
		\thanks{* Corresponding author: Jianzong Wang, \texttt{jzwang@188.com}.
		}
	}

	
	\maketitle
	
	\begin{abstract}
		The increasingly stringent regulations on privacy protection have sparked interest in federated learning. As a distributed machine learning framework, it bridges isolated data islands by training a global model over devices while keeping data localized. Specific to recommendation systems, many federated recommendation algorithms have been proposed to realize the privacy-preserving collaborative recommendation. However, several constraints remain largely unexplored. One big concern is how to ensure fairness between participants of federated learning, that is, to maintain the uniformity of recommendation performance across devices. On the other hand, due to data heterogeneity and limited networks, additional challenges occur in the convergence speed. To address these problems, in this paper, we first propose a personalized federated recommendation system training algorithm to improve the recommendation performance fairness. Then we adopt a clustering-based aggregation method to accelerate the training process. Combining the two components, we proposed Cali3F, a calibrated fast and fair federated recommendation framework. Cali3F not only addresses the convergence problem by a within-cluster parameter sharing approach but also significantly boosts fairness by calibrating local models with the global model. We demonstrate the performance of Cali3F across standard benchmark datasets and explore the efficacy in comparison to traditional aggregation approaches.
	\end{abstract}
	
	\begin{IEEEkeywords}
		Recommendation Systems, Federated Learning, Personalization, Convergence, Communication Cost
	\end{IEEEkeywords}
	
	\section{Introduction}
	As the most effective way to alleviate the information overloading dilemma, Recommendation Systems (RS) collect various characteristics of individuals such as demographic characteristics, rating information for items (explicit feedback) or users' interactions with specific items (implicit feedback) \cite{Bobadilla2013RecommenderSS} to generate recommendations. Conventional RS holds users' data centrally to facilitate the centralized model training, however, this increases privacy issues \cite{Zhang2018DiscreteDL}. Decentralized training frameworks arise swiftly as a result of rapidly rising privacy concerns. Google proposed Federated Learning (FL) \cite{Konecn2016FederatedLS,McMahan2017CommunicationEfficientLO} to help entities collaboratively train a global model through aggregating gradients to avoid the transmission of numerous sensitive data among devices, which has sparked a lot of interest in both academia and industry.

	In contrast to the traditional training method for centralized storage, federated learning is a machine learning setting that enables participating clients to solve a machine learning problem cooperatively under the coordination of a central server \cite{DBLP:journals/spm/LiSTS20}. The original data is not transferred or exchanged but stored locally during federated learning. Instead, after clients' activities, the server aggregates the gradients or modified model parameters on each round. Typical FL repeats five steps until the stop condition is met: client selection, broadcast current global model, client computation with local data, update aggregation and model updates. The original FL system deployed on Google's Gboard uses numerous users' smartphones as computational entities for collaborative training, which is known as cross-device FL architecture \cite{Kairouz2021AdvancesAO}. 

	Specific to recommendation algorithms, Collaborative Filtering (CF), which infers users' potential interests from historical behavior data, is at the heart of modern recommendation systems. Ammad et al. \cite{Ammaduddin2019FederatedCF} first proposed the FCF algorithm based on implicit feedback. 
	Whereafter the gradient leakage threat and interacted item list revealing problem are repulsed by several variant architectures \cite{Chai2021SecureFM, Minto2021StrongerPF, Liang2021FedRecLF, DBLP:conf/icassp/LiuZWX21}. However, enhancing FCF with deep neural networks has been largely unexplored. This is partly for practical reasons: significant improvements brought by most deep-learning models in RS can be seen only when numerous features of heterogeneous types are added to the input data \cite{Steck2021DeepLF}. However, in federated contexts, processing individual data in large-scale recommendation systems is difficult due to the natural heterogeneity of the data \cite{DBLP:conf/ijcnn/XueZWX21}. For this problem, we argue that it can be well solved using personalized models, and we propose a fast converging personalized model training framework to ensure that all clients get satisfactory model results.
	
	This work is motivated by the opportunity of combining recent advances in neural recommendation systems and FL. We first established a multi-task learning-based personalized recommendation model training framework to tackle the problem of heterogeneous data and ensure that the accuracies of recommendation advice for various users are relatively fair. After then, in order to avoid the additional training costs it brings, a quick update strategy was implemented to ensure that the user's overall computing overhead and communication costs are actually reduced.
	
	In a nutshell, the main contributions of this work are listed as follows:
	\begin{itemize}
		\item We combine deep learning with federated recommendation  system to improve recommendation performance in FL scenarios. To improve the performance fairness among the clients, we propose a training framework for personalized recommendation systems that uses global model parameters for calibration while personalizing.
		\item To compensate for the computational cost associated with personalization models, we use a update sharing method based on clustering, which could obtain satisfactory recommendation performance at the very early stage, thus effectively reducing the total training cost.
		\item Extensive experiments on benchmark datasets demonstrate the advantageous performance of our proposed architecture in terms of prediction accuracy, convergence speed and communication cost. 
	\end{itemize}
	
	\section{Related Work}
	\subsection{Federated Recommendation Systems}
	Matrix Factorization (MF) has acquired prevailing popularity as a latent factor model-based approach since the Netflix Prize. Later He et al. \cite{He2017NeuralCF} argued that MF's performance could be hampered by a too simple interaction function, and proposed Neural Collaborative Filtering (NCF), which improved performance by replacing the inner product with neural networks and using embedding layers to obtain the latent user-item vectors.
	
	To explore RS in the federated setting, Ammad et al. [3] first proposed a Federated Collaborative Filtering algorithm (FCF) based on implicit feedback. Their system simply uploads raw item profile updates, whereas user profile updates are calculated locally and aggregated through a series of non-reducible summations.
	
	Later, on the security side, Chai et al. \cite{Chai2021SecureFM} pointed out that if the gradient information of any two-step update and the update of user factor are given, the user rating information can be derived by solving higher-order equations. That is, gradient information may reveal user privacy data. To solve this problem, they proposed a secure matrix factorization model, SFMF, which eliminates the possibility of gradient leakage of user information by using the homomorphic encryption method in the communication process between server and client. On the generality side, the FedRec proposed by Lin et al. \cite{Lin2021FedRecFR} extends it to the explicit feedback recommendation scenario. During user gradient uploading, random samples of ungraded items are uploaded to the server to cover the actual item information exchanged by users, and user average rating and mixed rating mechanisms are adopted to generate negative sample item ratings. However, this approach introduced additional noise, so Liang et al. \cite{Liang2021FedRecLF} further proposed FedRec++ to eliminate noisy data in a privately-aware way by assigning part of the denoising clients, thus achieving better recommendation performance. Unfortunately, none of the aforementioned methods incorporates the benefits of deep networks into Federated Recommendation Systems (FedRS).
	
	When it comes to the convergence speed of FedRS, Muhammad et al. first introduced the clustering approach to FCF. By classifying users according to non-privacy profiles and adopting different update strategies to three model components, the FedFast \cite{Muhammad2020FedFastGB} algorithm could receive considerable practicability at the early stage of model training, but they only consider the simplest federated generalized MF algorithm and the resulting model is not fundamentally different from the one obtained by FedAvg \cite{Zhu2021FedRS}.
	
	\subsection{Performance Fairness in FL}
	As the most classical aggregation algorithm, FedAvg \cite{McMahan2017CommunicationEfficientLO} performs $E$ epochs of SGD on a randomly sampled fraction of devices and then averages their model updates according to instance numbers at each iteration. Although FedAvg has demonstrated its empirical success, it has been shown to encounter divergence in non-IID data distribution settings \cite{McMahan2017CommunicationEfficientLO,DBLP:conf/iclr/LiHYWZ20}.
	
	Since it is inevitable to face the challenge of data heterogeneity, participants have long been concerned about how to prevent the performance from significantly deviating across clients. The goal of performance fairness is to achieve a more uniform accuracy distribution across participants. It is worth mentioning that this applies to horizontal FL \cite{Yang2019FederatedML} where the participants are not in commercial competition with each other. 
	
	Several tactics are commonly adopted to encourage this kind of uniform accuracy distribution \cite{DBLP:conf/nips/0001MO20,DBLP:journals/expert/ChenQWYG20}. For instance, Li et al. \cite{Li2020FairRA} proposed a fair resource allocation method that improves the accuracy of the worst-performing devices through a novel optimization objective with a hyper-parameter to adjust the strength.
	Another framework, Agnostic Federated Learning (AFL) \cite{mohri2019agnostic}, treats the fairness as a min-max optimization problem and tends to minimize a worst-case objective.
	
	Closer to the essential point, given that data heterogeneity is the core cause of non-uniform performance in FL, a variety of techniques for personalized federated learning have been proposed. The Multi-task Learning Framework (MTL), which includes hard parameter sharing and the weighted combination method\cite{Zhang2021PersonalizedFL}, is one of the most popular personalization techniques. Smith et al. \cite{Smith2017FederatedML} first proposed a primal-dual MTL based personalized FL framework. Current mean-regularized MTL approaches, such as \cite{Dinh2020PersonalizedFL,Hanzely2020LowerBA} usually assign personalized models to clients and update them with a regularization term towards their current average. Ditto \cite{Li2021DittoFA}, on the contrary, simultaneously learn local and global models via a global-regularized MTL framework, where local models are regularized towards a global model trained by FedAvg rather than their average. The bi-level optimization proposed by Ditto is as follows:
	\begin{equation}
		\begin{aligned}
			&\min\limits_{v_k}\ h_k(v_k;w^*):=F_k(v_k)+\frac{\lambda}{2}\parallel v_k-w^*\parallel^2\\
			& s.t. w^*\in \mathop{\arg\min}\limits_w G(F_1(w),\dots,F_K(w))).
		\end{aligned}
	\end{equation}
	where the hyper-parameter $\lambda$ governs the interpolation between local and global models, finding a trade-off between robustness and fairness. $F(v_k)$, the function of local parameters $v_k$, denotes the local objective for client $k$ and $w^*$ denotes global parameters of the previous round. Plug-in style local models allow for methods designed for the global FL problem to be easily re-used in practice. However, training local models and the global model at the same time brings about a doubling of computing and communication costs, placing more demands on participants of FL. In addition to this, clustering algorithms based on clients' similarities \cite{DBLP:conf/ictc/YooSJJKYJC21,Sattler2021ClusteredFL,DBLP:conf/ijcnn/BriggsFA20} are leveraged to reduce non-IIDness of data within a single cluster although they allocate test set to all clusters or place additional requirements on the data domain.
	
	\section{Proposed Approach}
	In this section, we illustrate the Cali3F framework we proposed, which comprises three key components: clustering-based client sampling method (\textbf{ClusSamp}), calibrated local model update (\textbf{CaliUp}), and fast aggregation approach (\textbf{FastAgg}). We first introduce the base recommendation model we used in our experiments, then we explain in detail how our framework can be applied to the recommendation model and achieve advantages in convergence speed and recommendation fairness. 
	
	\begin{algorithm}[t]
		\caption{ Cali3F framework}
		\label{alg:algorithm1}
		\setstretch{1.25}
		\SetAlgoLined
		\KwIn{User set $U$; Item set $I$; Client representations $C$; Learning rate $\eta$; Personalized level $\varphi$; Cluster number $P$.}
		
		\BlankLine
		Initialize group $\mathcal{G}, D, S \leftarrow ClusSamp(C)$ \\

			\For{round t = 1,2,3,$\dots, T$ }{
				\ForEach{client $k$ in $D$ in parallel}{
					\texttt{/*~~Client side}\texttt{~~*/}\;
					$\triangle w^t_k\leftarrow \triangledown F_k(w^{t-1})$\\
					$\triangle v^t_k \leftarrow \triangledown F_k(v^{t-1}_p)$, k in cluster p
				}
				
				\texttt{/*~~Server side}\texttt{~~*/}\;
				
				\ForEach{cluster $p \in [P]$}{
					\texttt{\% calibrated update of non-embedding component}\\
					$\triangle v_p^t = \sum \limits_{k\in D_p}\frac{n_k}{n_{\sigma}},n_{\sigma}=\sum\limits_{k\in D_p} n_k$\\
					$v_p^t[N] = v_p^{t-1}[N]-\eta \left( \triangle v_p^t+\varphi \parallel \triangle v_p^t \parallel \frac{v_p^{t-1}-w^{t-1}}{\parallel v_p^{t-1}-w^{t-1}\parallel} \right)$
				}
				\texttt{\% FedAvg manner}\\
				$w^t[N] = w^{t-1}[N]-\eta \sum\limits_{k\in D}\triangle w_k^t[N],n_{\sigma}=\sum\limits_{k\in D} n_k$\\
				\texttt{\% update user and item embeddings}\\
				$w^t[U,I] \leftarrow FastAgg(\{\triangle v_k^t\}_{k\in D})$
				
				\texttt{\% re-clustering and average local models within clusters}\\
				$\mathcal{G}, D, S \leftarrow ClusSamp(w_t[U])$\\
				$\{v_p^t\}_{p\in [P]} \leftarrow ClusAvg(\{v_p^t\}_{p\in [P]},\{n_k\}_{k\in [K]}) $
			}			
			\KwOut{Personalized recommendation models $\bm{\{v_p\}_{p\in[P]}}$; Global model $w$} 
		\end{algorithm}
		Along with the coming of deep neural networks' prevailing, the complex models are likely to generate more accurate predictions, but with more parameters to learn, thus imposing communication and computation costs on the clients. However, the clients in cross-device federated learning are usually thin (e.g. mobile phones), which propose a challenge to the trade-off between recommendation accuracy and limited bandwidth and computing resource. We tend to implement a resource-friendly deep model to meet the constraint for thin devices. The best choice would undoubtedly be NCF \cite{He2017NeuralCF} with the advantages of low latency and low cost. We divide the model $w$ into three parts $w[U,I,N]$ according to their functions: \textbf{user embedding component}, \textbf{item embedding component} and \textbf{non-embedding component} respectively.
		\subsection{Personalized Federated Recommendation}
		
		In this paper, we develop a scalable personalization technique to explore how high averaged accuracy and better fairness amongst clients can coexist. As we will see, this lightweight personalization framework can easily enhance existing recommendation algorithms while still having a good empirical performance by incorporating federated multi-task learning. 
		To clarify the objectives of the study, we first present a formal definition of fairness in a horizontal FL setting following Li et al. \cite{Li2020FairRA}:
		\newtheorem{definition}{Definition}
		\begin{definition}
			\label{def}
			(Fairness). For two models $w_1$ and $w_2$, we say that $w_1$ is more \textit{fair} than $w_2$ if the test performance of model $w_1$ on $K$ clients is more \textit{uniform} than that of $w_2$'s, \textit{i.e.}, $std\left(\{NDCG(w_1)_k\}_{k\in[K]}\right)<std\left(\{NDCG(w_2)_k\}_{k\in[K]}\right)$ where $NDCG(\cdot)_k$ denotes the normalized discounted cumulative gain on device $k$ and $std(\cdot)$ denotes the standard deviation.
		\end{definition}
		In general, we consider two 'tasks': the global objective $F_k(w)$, and the local objective $F_k(v_p)$, which aims to learn a model using only the input from device $k$. To connect global and local tasks, we encourage personalized models to be close to the concurrently optimized global model state. As shown in Figure \ref{fig:grad}, we use a simple yet effective additional regularized loss function with an adaptive $l_2$ norm for each client.
		\begin{equation}
			\frac{\varphi}{2}\frac{\triangledown F_p(v_p)}{\parallel v^p-w\parallel} \parallel v_p-w \parallel ^2
		\end{equation}

		Here $\triangledown F_p(v_p)$ denotes the average of delegates' gradients. The interpolation between local and global models is controlled by the hyper-parameter $\varphi$. When $\varphi$ is set to 0, the personalized process is reduced to training local models independently; as $\varphi$ grows larger, it is closer to the global model.
		
		\subsection{Clustering-based Sampling Method (ClusSamp)}
		We emphasize that only using the aforementioned personalized technique will result in improved fairness and robustness, as well as a doubling local calculation cost, due to the repetitive optimization procedure. In fact, we argue that for recommendation tasks, changing the local model client-by-client is unnecessary. Based on the user relevance employed by the recommendation system, we can assume that similar clients benefit from similar recommendation models. To put it another way, we might be able to prevent diversions and arrive at the most harmonic solution if we sample more representative users per round when updating the federated model.
		
		In this case, we adopt ClusSamp algorithm.  First and foremost, ClusSamp groups clients into $\mathcal{G}$ of $P$ clusters based on clients' privacy-neutral metadata that guarantee their anonymity. Then ClusSamp selects a given number $m/P$ of clients from per cluster at random to participate in FL. The clients selected in this round are referred to as delegate clients (delegates), while the remaining are called subordinate clients (subordinates). We designate delegates and subordinates in cluster $p$ as $D_p\subset D$ and $S_p \subset S$, respectively, in the next parts of the method.
		
		This method of initializing the FL approach allows us to select a representative sample of the domain's various user populations. In practice, it may be more feasible to cluster the clients in $[K]$ using additional privacy-preserving criteria (e.g., region, device type) to produce clusters that reflect the ideal user groups in $[K]$. Nonetheless, due to the limitation of existing datasets, we are not currently using additional features, which proves to do no damage to this algorithm's effectiveness. 
		
		\subsection{Fast Aggregation Benefits from Delegates (FastAgg)}
		\label{sec:aggregation}
		
		There is an opportunity to improve the efficiency of this aggregation procedure. Parameter updates learned by any user A can be applied in some way to a similar user B, thus accelerating the learning process for user B and consequently improving the federated system's overall efficiency. By updating $\mathcal{G}$ at the end of each round, we can maintain accurate clusters of clients with similar user embeddings and broadcast client updates to all other clients of the cluster that contains them. FastAgg's aggregation approach boils down to this. FastAgg combines local models to update the shared global model in a manner that causes the federated model to converge faster and produce higher-quality recommendations. 
		
		FastAgg works by embedding calibration and sharing the training progress of delegates with their subordinates. This sharing is especially useful for speeding up learning in the early rounds, but it becomes less significant as the model converges. As a result, we use a discounting factor $\lambda$ to manage its impact. 
		With this server-based intra-group sharing of training experiences, client workloads and reduce communication rounds to the server should be minimized, and recommendation performance and model convergence speed should be improved.
		
		\noindent\textbf{Update of delegate user embeddings.}
		We can easily update delegates' embedding with their actual updates because we separate clients into various clusters and the user-sets do not overlap:
		\begin{equation}
			w^t[U_d]\leftarrow w_d^t[U_d]\ \textit{ for each }d \in D
		\end{equation}
		where $w[U_d]$ denotes the part of user embedding parameters of the delegate user $d$ while $U_d$ refers to its indice.
		
		\noindent\textbf{Update of subordinate user embeddings.}
		We update the subordinate user embeddings with the averaged update of the corresponding delegate user embeddings, assuming that delegates belonging to each cluster can represent their subordinate users at each round. To avoid network fluctuation during late training, a discount factor $\lambda$ is required. The updates of subordinate user embeddings in cluster $p$  are given by:
		\begin{equation}
			\begin{aligned}
				w^t[U_s] \leftarrow w^{t-1}[U_s]&+\lambda * mean( \triangle w[U_d])\\
				&\textit{  for each } s \in S_p, d \in D_p
			\end{aligned}		
		\end{equation}
		
		\noindent\textbf{Update of item embeddings.}
		Instead of being associated with the relevant delegate user client, the same item may have interacted with other users, causing all users who interacted to update the item vector at the same time when they are chosen to participate in the training. Items that never appear in this round will lower the size of the single update if we simply average the item embedding matrix. So $w[i]$, the item embedding parameters corresponding to item $i$, is updated by a weighted average of $w^t[i]$ learned by delegates in each round of updates to make the most of all the interactive information. The update quantity is used to calculate each delegate's share:
		\begin{equation}
			\begin{aligned}
				&I_d = \{i \in I\ |\ w^{t}[i]-w^{t-1}[i]\neq0\} \text{, for } d \in D  \\[2.5mm]
				& \theta^d[I_d]= |w^{t}_d[I_d]-w^{t-1}[I_d]| \text{, for } d \in D ,\\[1.5mm]
				& w^t[i] = \frac{1}{\sum \limits_{d\in D} \theta^d[i]} \sum\limits_{d\in D}\theta^d[i]w^t_d[i] \text{, for }i \in I 
			\end{aligned}
		\end{equation}
		In this way, multiple clients can contribute to the update of the item vector.
		
		\noindent\textbf{Update of non-embedding component (CaliUp).}
		Personalization is widely-studied in FL. The hidden layer closest to the output layer in neural networks usually contains the most personalized information. Instead of maintaining a local model that varies from place to place, we regard the non-embedding layer as the personalization layer to capture personal information about the devices. 
		Cali3F's premise is that by reducing reliance on the global model, we may reduce the representation disparity and thus promote fairness. 
		
		We propose a method to calibrate the local model's training with the global model while keeping the differences from the global model. At each FL round, each cluster maintains a local model $v^p$. Except for the non-embedding component, the other component of $v^p$ follows the global model. For brevity's sake, we have omitted the sign $``[N]"$ that follows the model parameter signs in the following formula.
		\begin{figure}[!htbp]
			\centerline{\includegraphics[width=7cm]{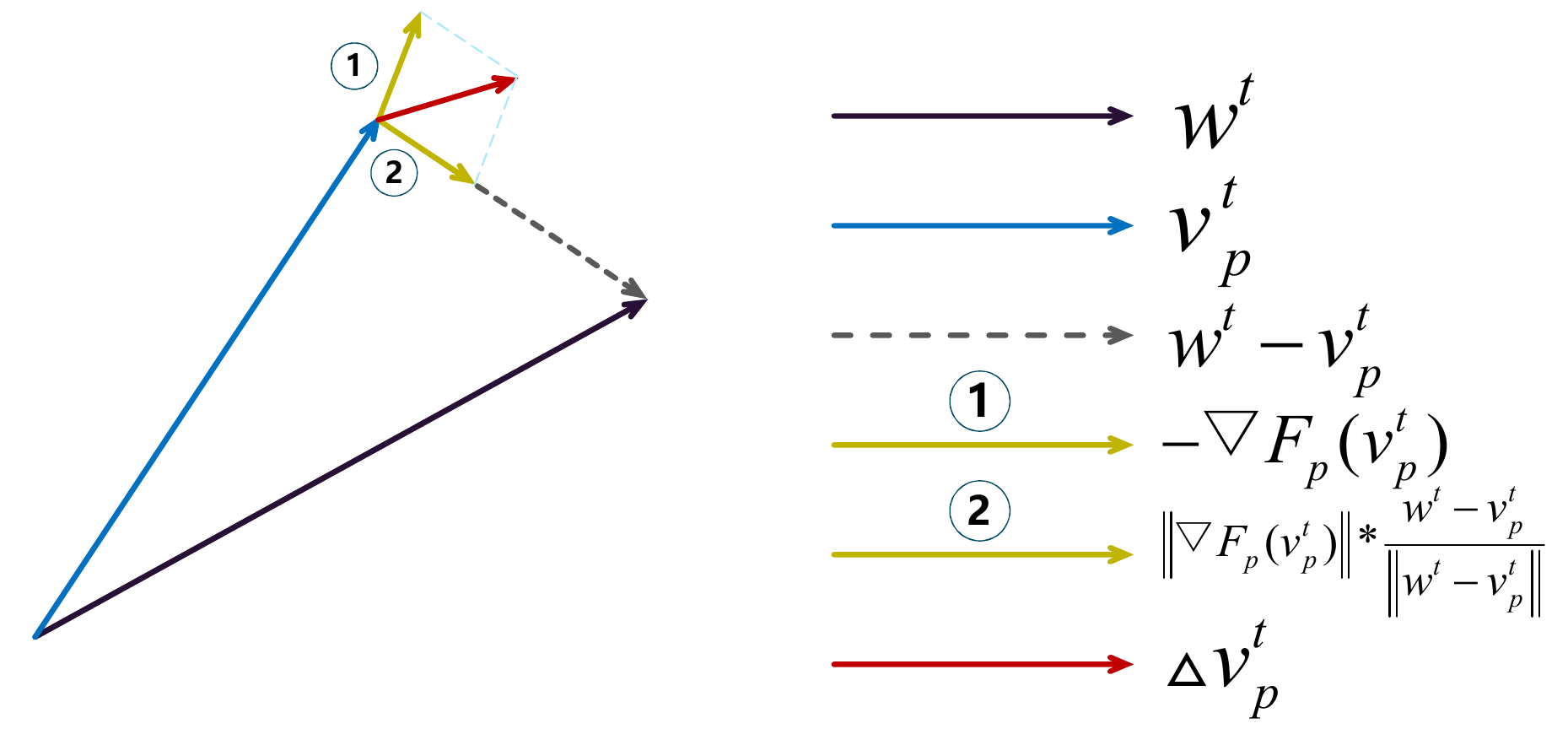}}
			\caption{Illustration of local model's updating process}
			\label{fig:grad}
		\end{figure}
		
		As depicted in Figure \ref{fig:grad}, we first calculate $_p$'s gradient using data from delegates that share this cluster. Then we apply an extra gradient, whose direction is towards the difference between the global non-embedding component and the local non-embedding component, while the norm is a fraction of the local gradient, to place a constraint on the local model that prohibits it from diverging too far.
		\begin{equation}
			\resizebox{.91\hsize}{!}{$v_p^{t+1}\leftarrow v_p^t - \eta \left(\triangledown F_p(v_p^t) + \varphi \parallel \triangledown F_p(v_p^t)\parallel \frac{v_p^t-w^t}{\parallel v_p^t-w^t \parallel}  \right)$}
		\end{equation}
		
		\subsection{Re-partitioning and Averaging within Clusters}
\begin{figure}[htbp]
			\centerline{\includegraphics[width=9cm]{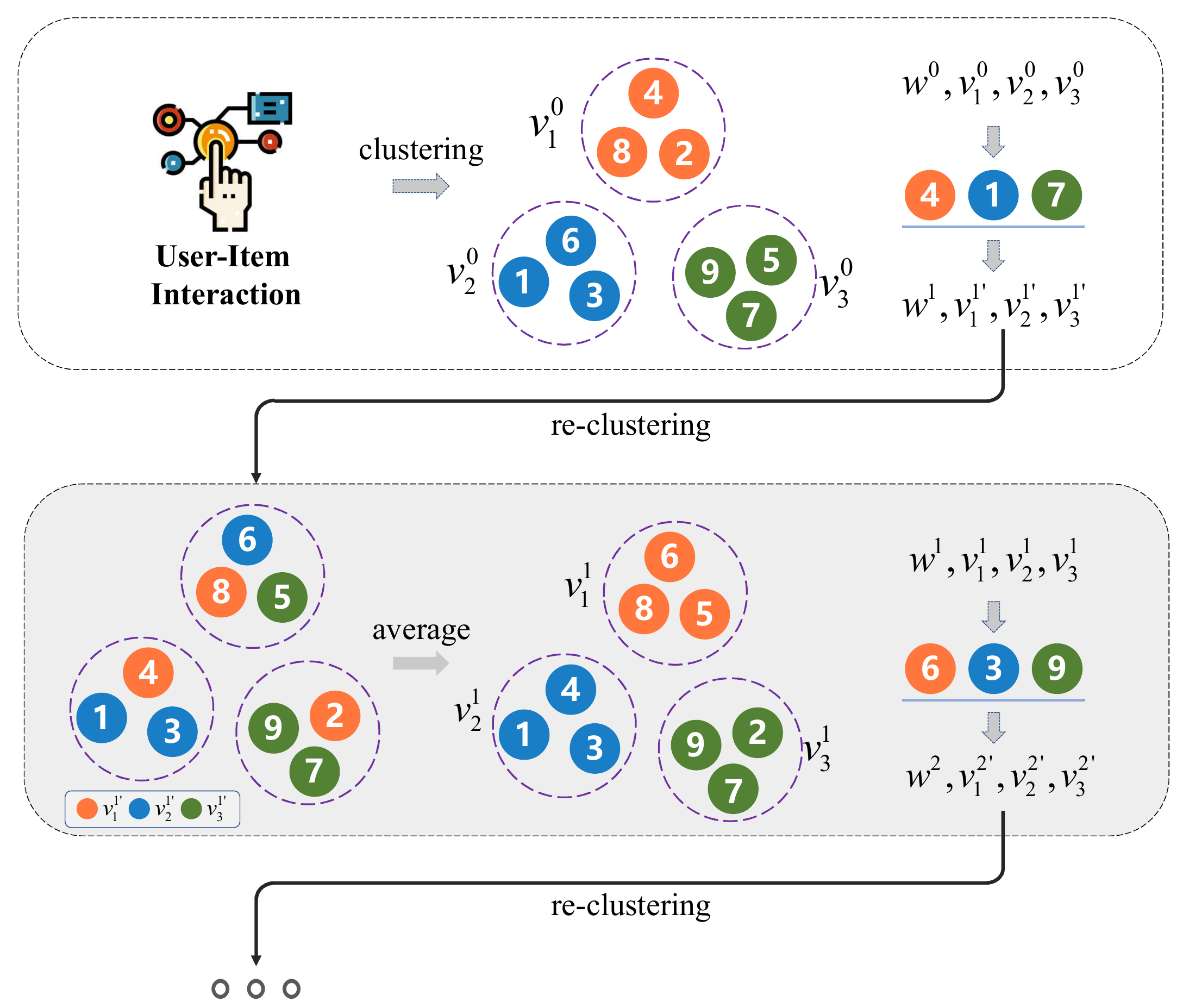}}
			\caption{Averaging of parameters across clusters. ClusSamp samples one client per cluster at random in a round-robin manner until the fraction of sampled clients meets the criteria at the end of each round. The clients are reassigned after training, and the non-embedding components of local models are weighted averaged based on the current clients in each cluster.}
			\label{fig:architecture}
		\end{figure}

		The user embeddings have changed to varying degrees after one cycle of training, as seen in Figure \ref{fig:architecture}. As a result, we assign a new client partition, $\mathcal{G}$, based on the newly obtained embeddings. Specifically, here we employ the K-means clustering method to re-cluster users into $P$ clusters. Clients who share the same cluster then average their non-embedding component parameters with respect to instance numbers to create a new local model that represents the cluster. Remarkably the final re-partitioned local models will be used for inference.
		\vspace{-0.31cm}
		\section{Experiments}
		In this section, we conduct the experiments that aim to answer the following research questions (RQ):\\
		\textbf{RQ1} Can our proposed federated version NCF gain considerable recommendation performance?\\
		\textbf{RQ2 }Does Cali3F improve fairness between clients?\\
		\textbf{RQ3} Does Cali3F converge faster than baseline FedRS algorithms?\\
		\textbf{RQ4 }How sensitive is Cali3F to its hyper-parameters in terms of convergence speed and fairness?\\
		\begin{figure*}
			\centering 
			\includegraphics[height=0.38cm,width=2.6cm]{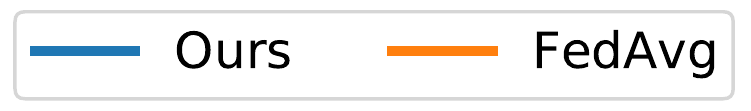}\\
			\centering
			\subfigure[FedGMF]{
				\includegraphics[height=3cm,width=5cm]{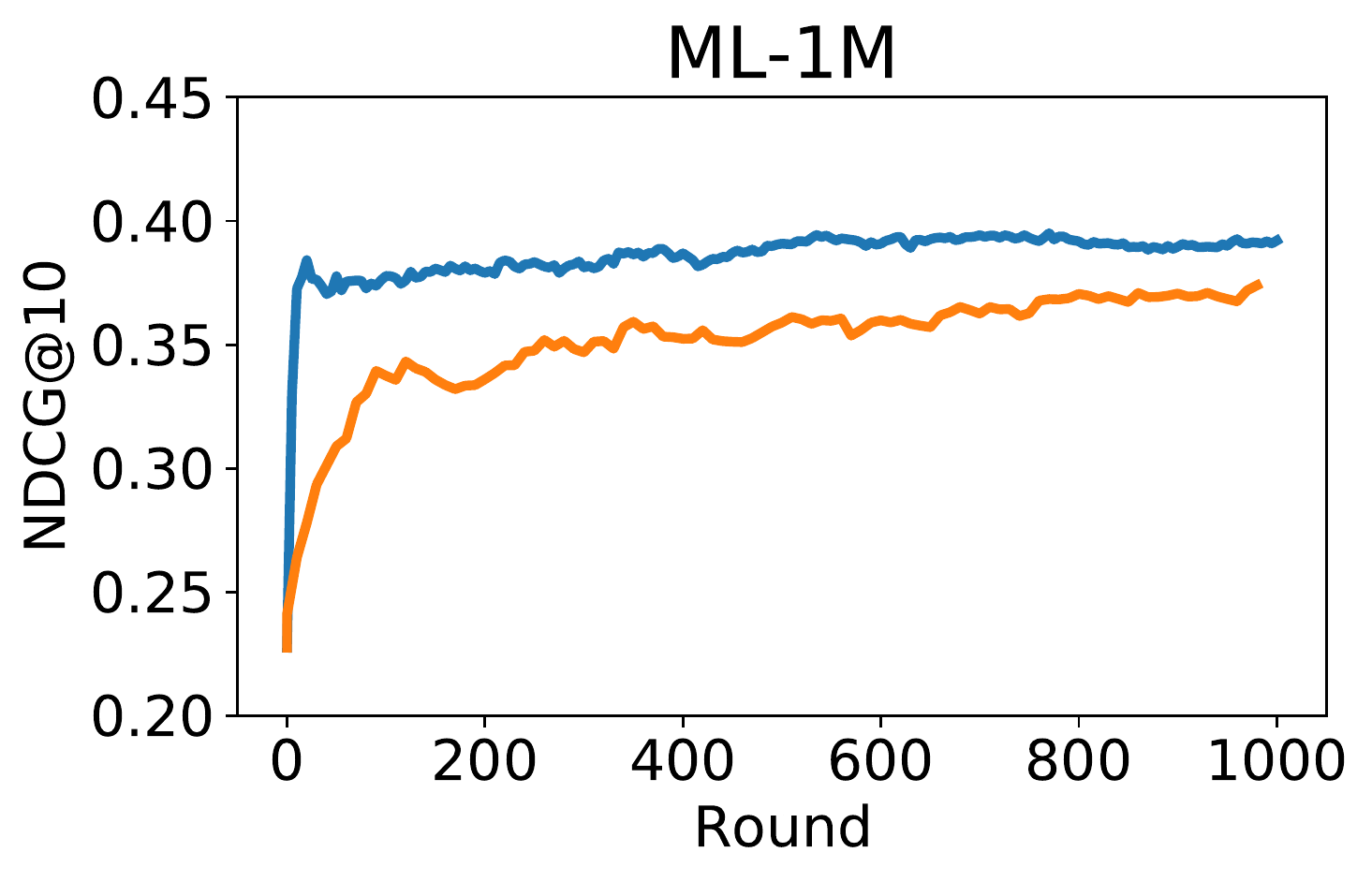} 
				\includegraphics[height=3cm,width=5cm]{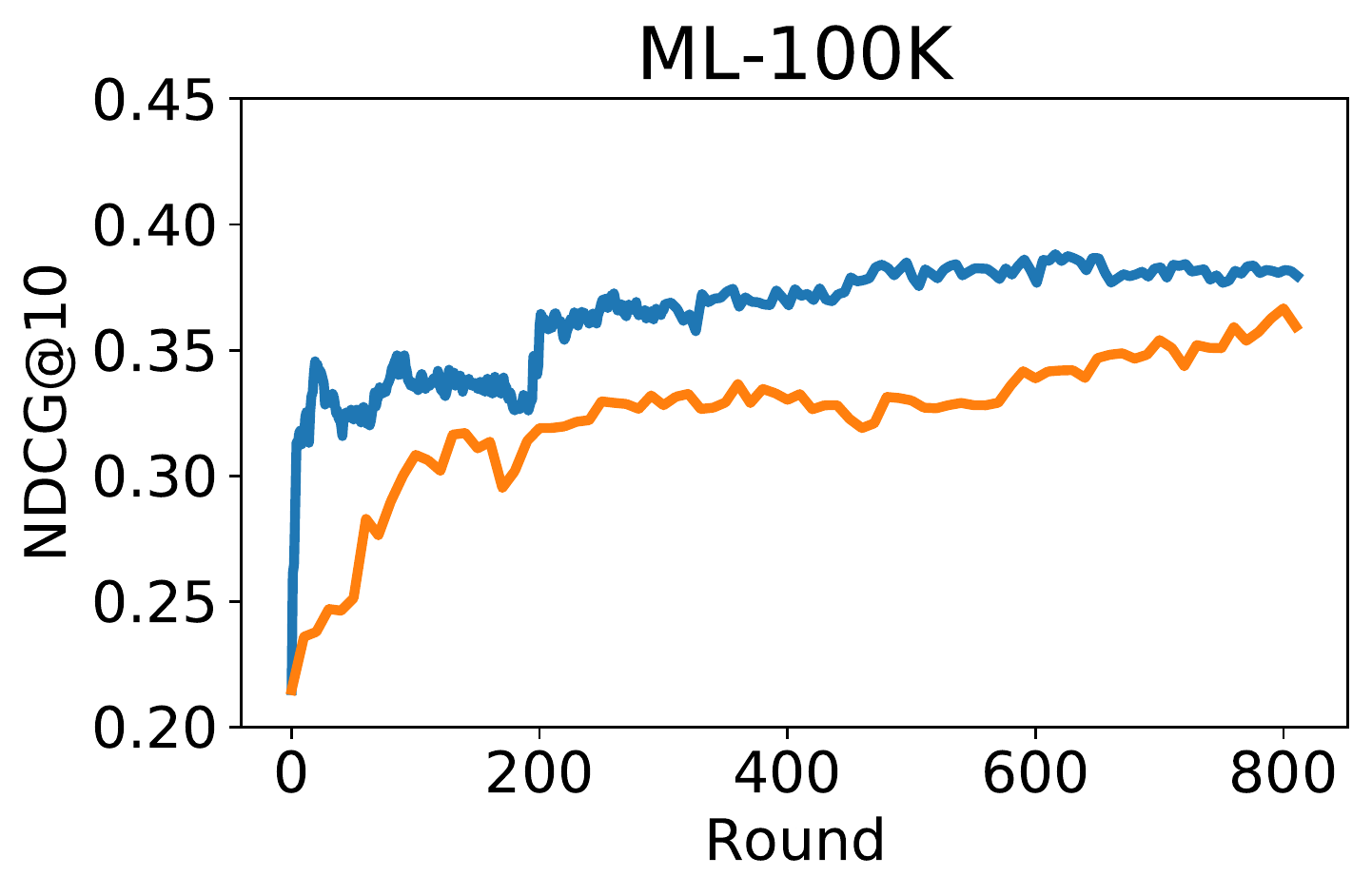} 
				\includegraphics[height=3cm,width=5cm]{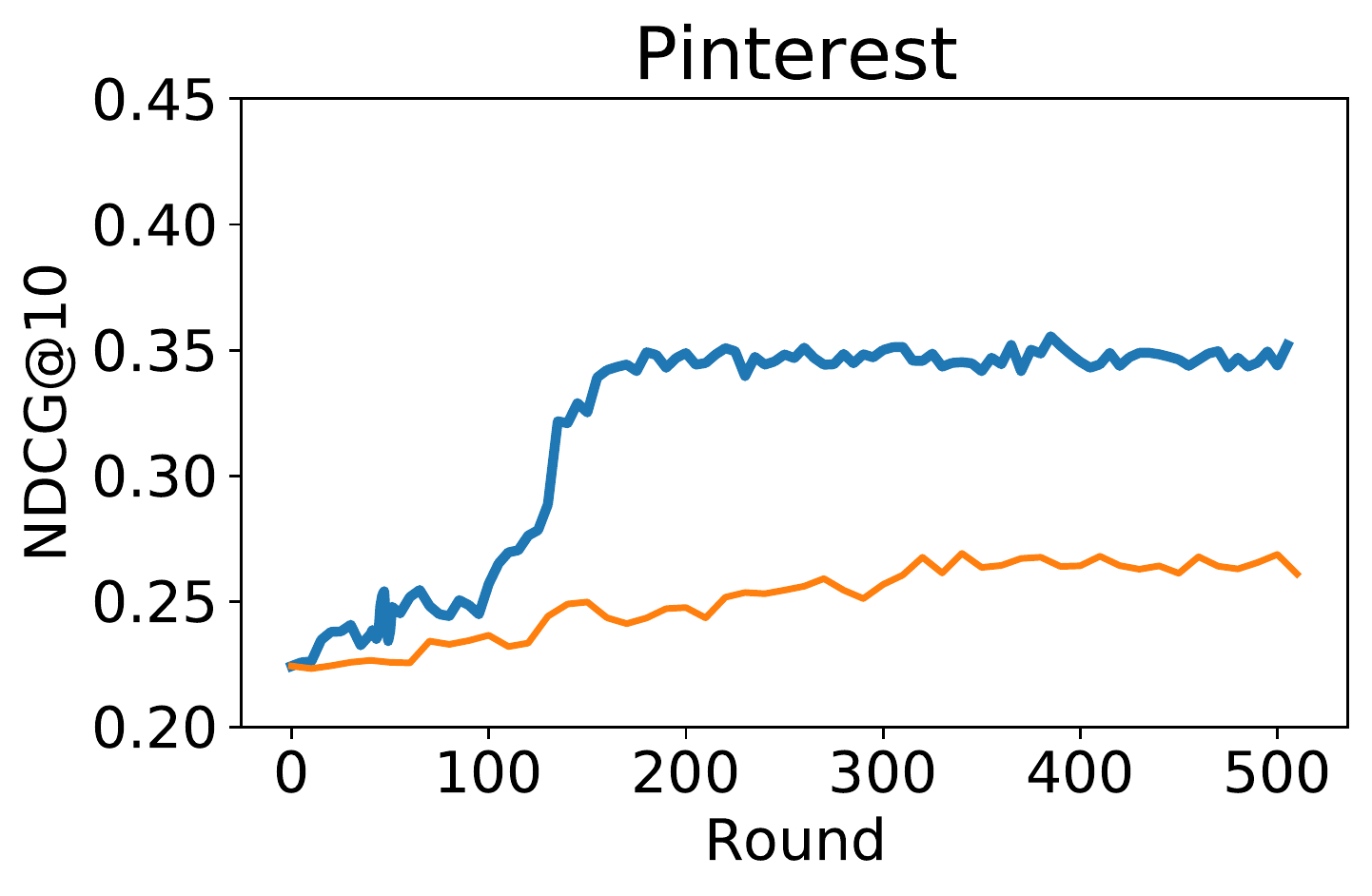}
			}\vspace{-0.2cm}
			\centering
			\subfigure[FedMLP]{
				\includegraphics[height=3cm,width=5cm]{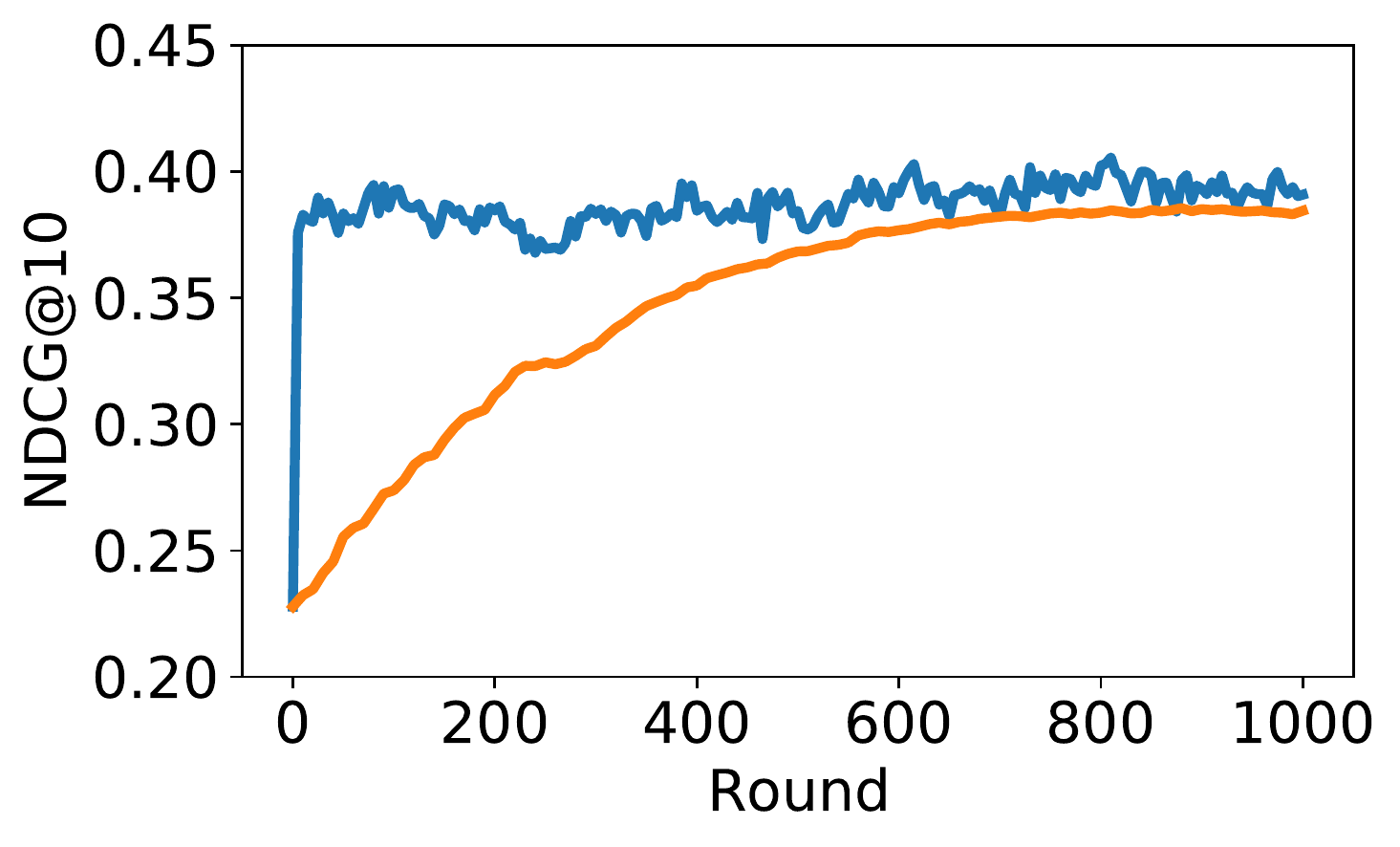} 
				\includegraphics[height=3cm,width=5cm]{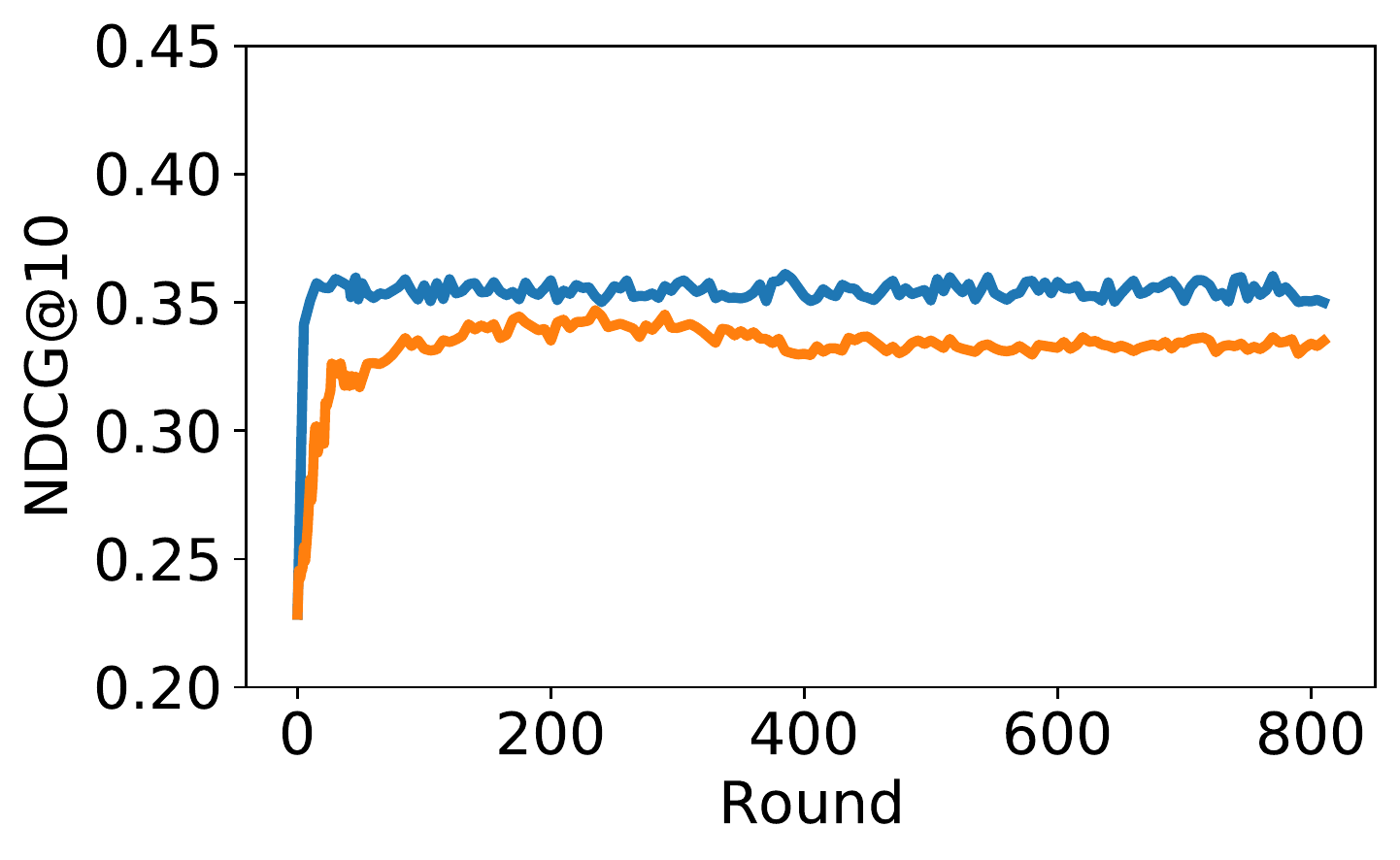}
				\includegraphics[height=3cm,width=5cm]{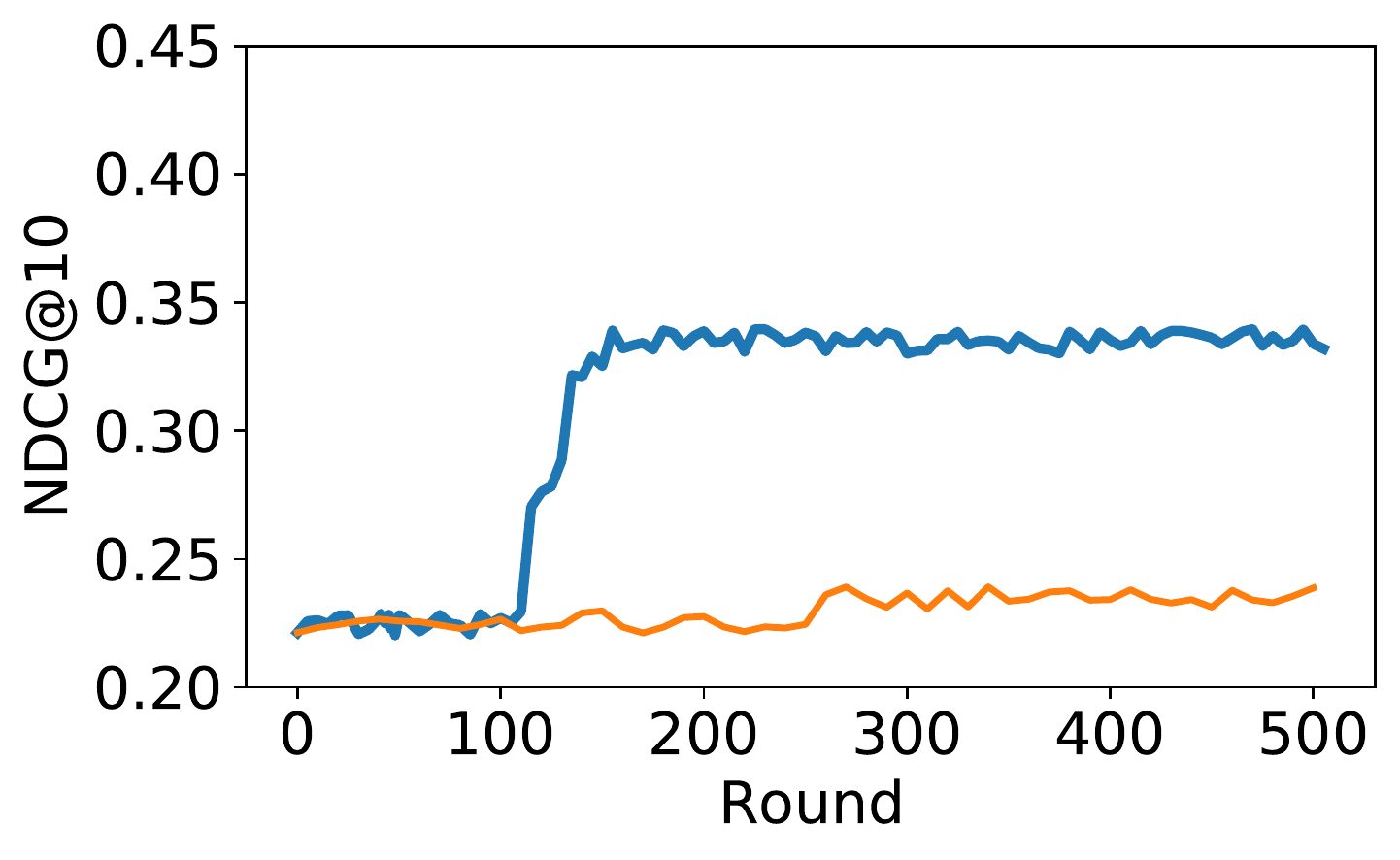}
			}\vspace{-0.2cm}
			\centering
			\subfigure[FedNeuMF]{
				\includegraphics[height=3cm,width=5cm]{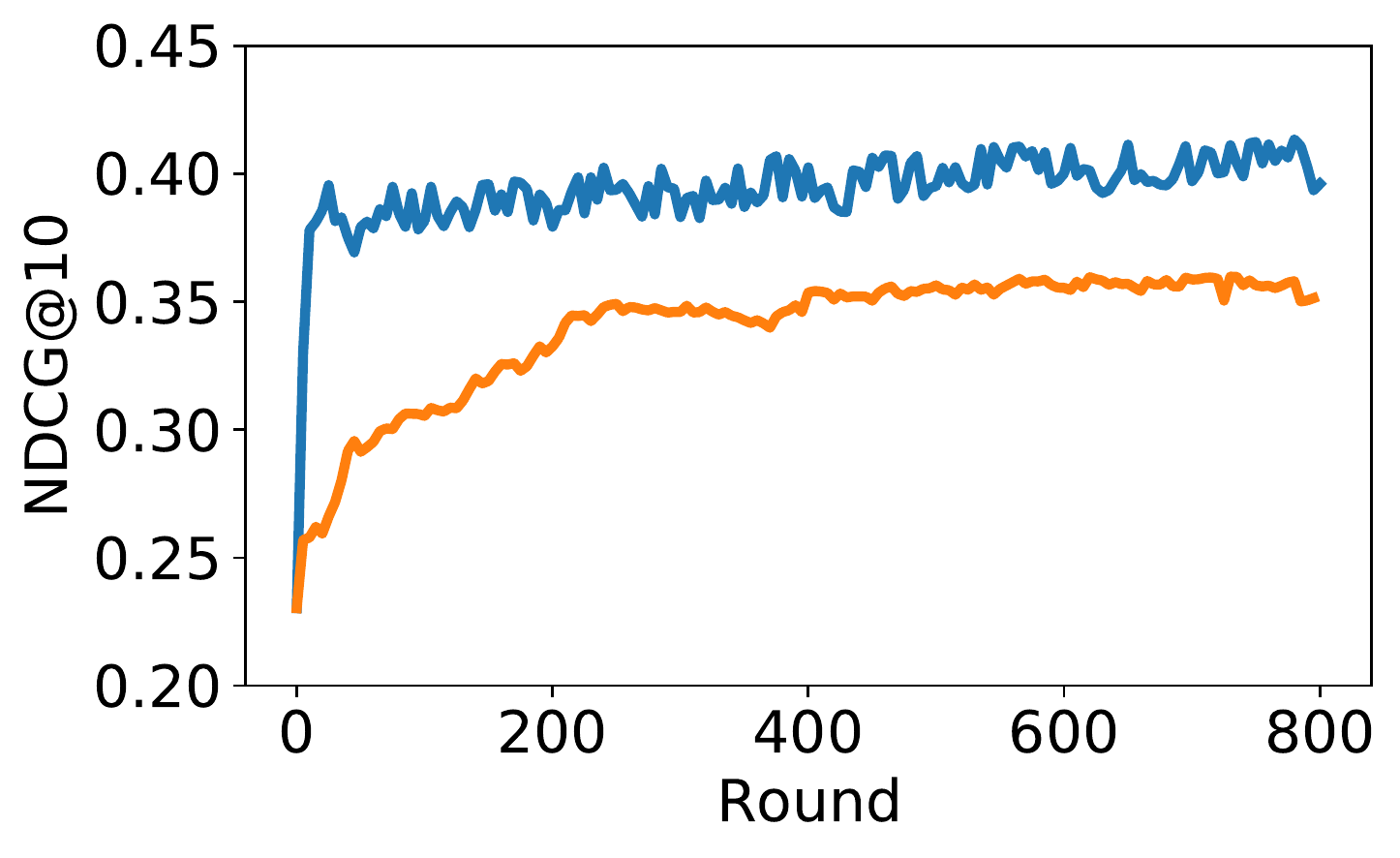}
				\includegraphics[height=3cm,width=5cm]{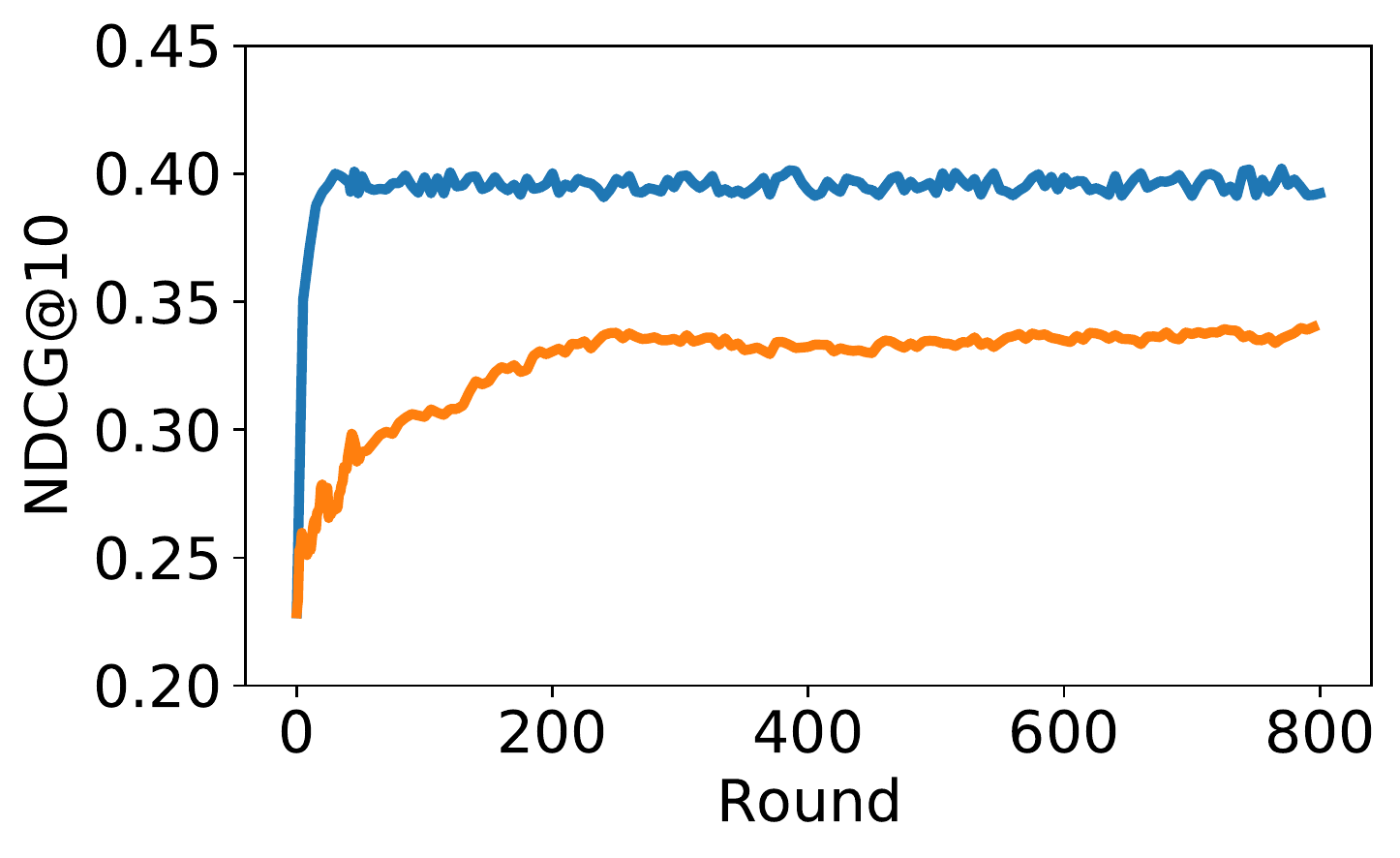} 
				\includegraphics[height=3cm,width=5cm]{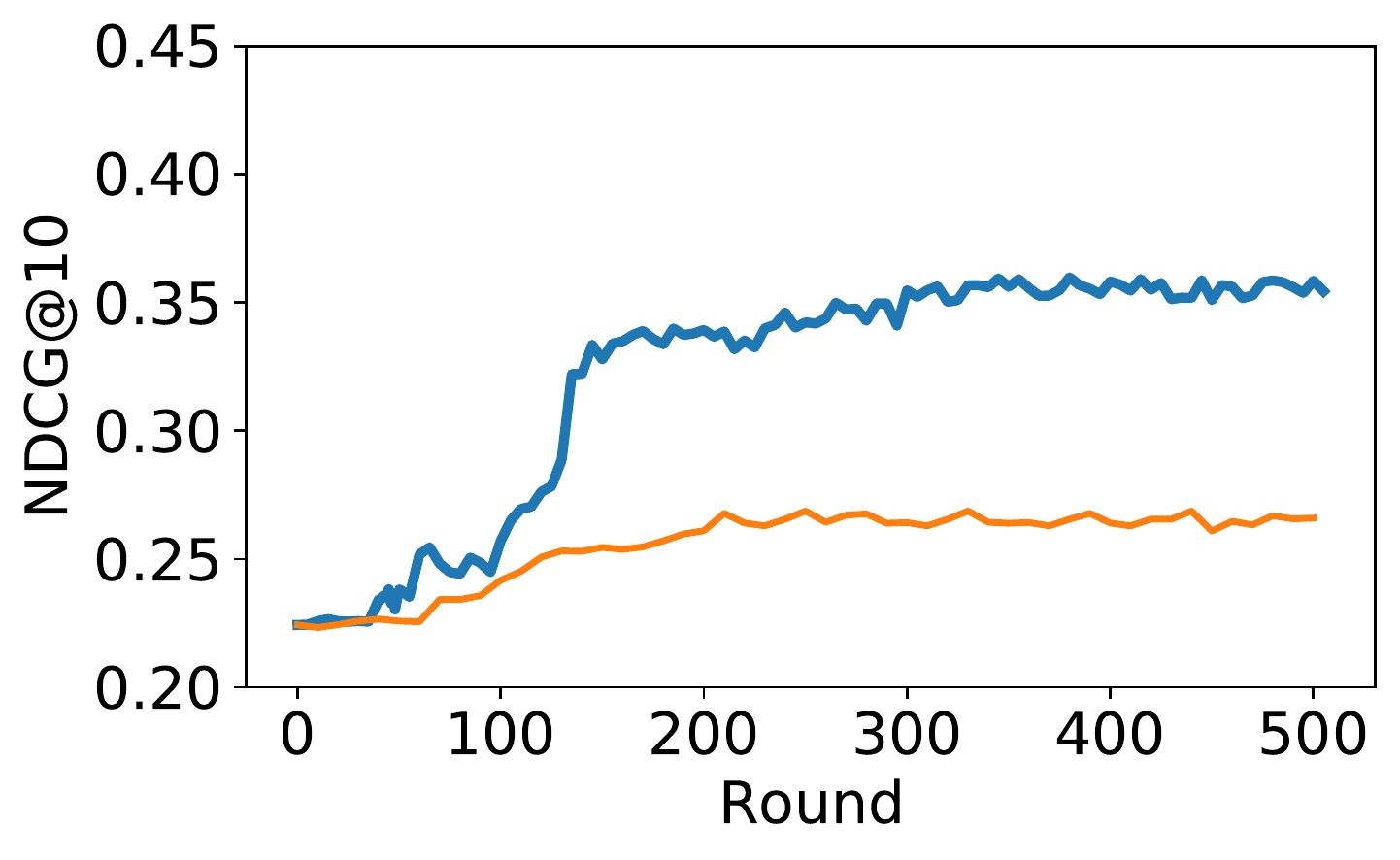}
			}
			\caption{Convergence speed comparison between different training settings. Number of clusters and delegates each cluster are fixed to guarantee Cali3F samples the same number of clients as FedAvg.}
			\label{fig:convergence}
		\end{figure*}\vspace{-0.2cm}
		
		\subsection{Datasets Descriptions}
		We test our proposed architecture on three classical publicly accessible datasets: MovieLens 1M\footnote{https://grouplens.org/datasets/movielens/1m/}, MovieLens 100K\footnote{http://grouplens.org/datasets/movielens/100k/} and Pinterest\footnote{https://sites.google.com/site/xueatalphabeta/academic-projects}, which are all benchmark datasets for collaborative filtering algorithm comparison. Because the first two only contain users with at least 5 interaction records, we simply transform them into implicit data, where each entry is tagged 1 if the user has rated the item, else 0. Pinterest is constructed for evaluating content-based image recommendation.
		It is implicit data but highly sparse, so we filtered users with less than 20 interactions. Table \ref{tab:datasets} shows the statistical description of all datasets after preprocessing. 
		\footnoterule
		\begin{table}
			\centering
			\fontsize{9}{13} 
			\selectfont
			\caption{Dataset statistics}
			\label{tab:datasets}
			\begin{tabular}{ccccc}
				
				\toprule
				Dataset & \#Interaction & \#User & \#Item & Sparsity \\
				\midrule
				\textbf{MovieLens 1M} & 1,000,209 & 6040 & 3706 & 95.53\% \\
				\textbf{MovieLens 100K}	& 100,000 & 943 & 1682 & 93.7\% \\
				\textbf{Pinterest}& 1,500,809 & 55,187 &9,916 & 99.73\% \\
				\bottomrule
			\end{tabular}
		\end{table}
		
		\subsection{Evaluation Metrics}
		We adopted the \textit{leave-one-out} evaluation to evaluate the performance of next item recommendation. For each user, we utilized her all interacted items for training, with the exception of the last interaction item, which was used as the test item. The ranked list includes the test item as well as 100 non-interacted items chosen at random. The performance of the recommendation is then assessed using \textit{Hit Ratio} (HR) and \textit{Normalized Discounted Cumulative Gain} (NDCG), both of which are calculated at 10 and averaged across all test users. HR calculates the percentage of test items that appear in the top 10 of the recommended list while NDCG divides the relevance results by $\log_2(i+1)$ to encourage higher rankings of test items, where $i$ is the position occupied by the test item, and then the sum is divided by the optimal result to normalize. 
		For the fairness of the recommendation model, we calculate the above metrics for the local model on local data, and then use the standard deviation to characterize the degree of fairness as mentioned in Definition \ref{def}. The larger the standard deviation indicates that the performance gap between the client models is larger and less fair. 
		As for the training speed, we experiment on Intel Xeon Gold 6130 CPU@2.10Hz and 16G Tesla V100 GPU to simulate the time cost. The code is constructed on the basis of TensorFlow.
		
		\subsection{Baselines}
		To demonstrate the benefits of our method in terms of convergence speed, we compared it to the widely used FedAvg algorithm. On three different datasets, we practiced the training process of three different NCF variants respectively. For convergence comparison, We picked NDCG, which has larger volatility for line drawing, as the measuring index to more clearly demonstrate the boundary between the two methods. For our original intention, fairness, we compared Cali3F with another method Ditto, which uses the $l_2$ norm of the difference between local and global models' parameters as a constraint term to train local models, but varies from our method on the fixed coefficient of regularization term. Since Ditto is a training method attached to the FedAvg training process, which does not do much to accelerate convergence, it will increase the amount of computation. Obviously. There is no comparability in the convergence speed. We only compare with Ditto and FedAvg on the improvement in fairness and reduction in the training time. 
		\subsection{Results and Discussion}
		\subsubsection{Recommendation Performance Comparison (RQ1)}
		Table \ref{tab:perf} shows the performance of HR@10 and NDCG@10 with respect to different training methods. We adopt the same hyper-parameter setting for NCF-based methods with two kinds of training methods, including the same number of latent factors, learning rate, number of MLP layers, etc. For MF methods, we refer to the results from \cite{He2017NeuralCF, PERIFANIS2022108441}. 
		\begin{table}[!htbp]
			\setlength{\tabcolsep}{1.6mm}{ 
				\centering
				\fontsize{9}{12}
				\selectfont
				\caption{Recommendation performance for three versions of Cali3F and traditional federated methods. The last four columns are centrally-trained baselines. Highest \textit{HR@10} and \textit{NDCG@10} observed are listed below.}
				\scalebox{1}{ 
					\begin{tabular}{lcccccc}
						\toprule
						&\multicolumn{2}{c}{\textbf{ML-1M}} & \multicolumn{2}{c}{\textbf{ML-100K}} & \multicolumn{2}{c}{\textbf{Pinterest}}\\
						& HR & NDCG & HR & NDCG & HR & NDCG\\
						\midrule
						GMF&0.69&0.41&0.68&0.42&0.87&0.56\\
						MLP & 0.63 & 0.36 & 0.61 & 0.35 & 0.85 & 0.54\\
						NeuMF & 0.70 & 0.41 & 0.69 & 0.42 & 0.88 & 0.55\\
						BPR & 0.66 & 0.40 & 0.92 & 0.30 & 0.87 & 0.54\\
						\midrule
						FedAvg-NeuMF&0.60          &0.37           &0.70          &0.36          & 0.50          & 0.24\\
						Cali3F-GMF&\textbf{0.73} &0.40           &0.70          &0.38          & 0.69          & 0.35\\
						Cali3F-MLP&0.72          &0.40           &0.69          &0.36          & 0.71          & 0.34\\
						Cali3F-NeuMF&0.72          &\textbf{0.41 } &\textbf{0.71} &\textbf{0.40} & \textbf{0.73} & \textbf{0.36}\\
						\bottomrule
					\end{tabular}
				}
				\label{tab:perf}
			}
		\end{table}
		It is worth mentioning that we are not attempting to tune a new state-of-the-art recommendation performance, which is a mission better suited to much more complicated neural networks. Rather, to demonstrate the universality of our framework modified on various methods. In comparison with federated methods, our Cali3F consistently outperforms the FedAvg baseline. When compared to respective upper bound with centralized training settings, the results show that using locally stored data is invariably accompanied by a decline in recommendation quality, which has been demonstrated in a wide range of literature and is not an impediment to showing the superiority of our approach.
		
		\subsubsection{Convergence Analyses (RQ2\&RQ3)}
		In subsection \ref{sec:aggregation}, we construct a rapid aggregation mechanism to compensate for the double local computation cost generated by the local-global model training algorithm. To validate the improvement effect of our aggregation strategy, we conduct a comparison on convergence speed. The idea behind it is that if significantly faster convergence can be accomplished, the cost savings on overall computing remain self-evident. The results are shown in Figure \ref{fig:convergence}. To be compared fairly, both methods start with the same initial model.
		
		Surprisingly, it is immediately obvious that Cali3F converges towards its best values even at the very early stage. Take ML-1M for instance, Cali3F settles to within 5\% of its best values at round 10, which means Cali3F can omit 98.6\% of training rounds while saving the same percentage in communication costs.
		
		\subsubsection{Ablation Study (RQ2\&RQ4)}
		\label{sec:ablation}		
		Cali3F's role in enhancing recommendation performance fairness amongst clients is investigated in this experiment. In Table \ref{tab:std}, We first report the respective fairness metric values on the most sophisticated NeuMF model, as well as the time cost to reach within 5\% of the best NDCG@10. The time cost is expressed as multiples of how long FedAvg takes. Cali3F's high efficiency and its promotion for fairness may easily be seen in Table \ref{tab:std}.
		\begin{table}
			\centering
			\fontsize{9}{12}
			\selectfont
			\caption{Fairness and time-cost comparison on FedNeuMF on ML-1M}
			\label{tab:std}
			\begin{tabular}{lccc}
				\toprule
				& FedAvg & Ditto & Cali3F \\
				\midrule
				Std &0.38  &0.33 & 0.33 \\
				Time-cost & 10.6min(1.00$\times)$ & 34.5min(3.25$\times$)  & 3.4min(0.32$\times$) \\
				\bottomrule
			\end{tabular}
		\end{table}
		
		
		\begin{figure}[!htbp]
			\centerline{\includegraphics[width=8cm]{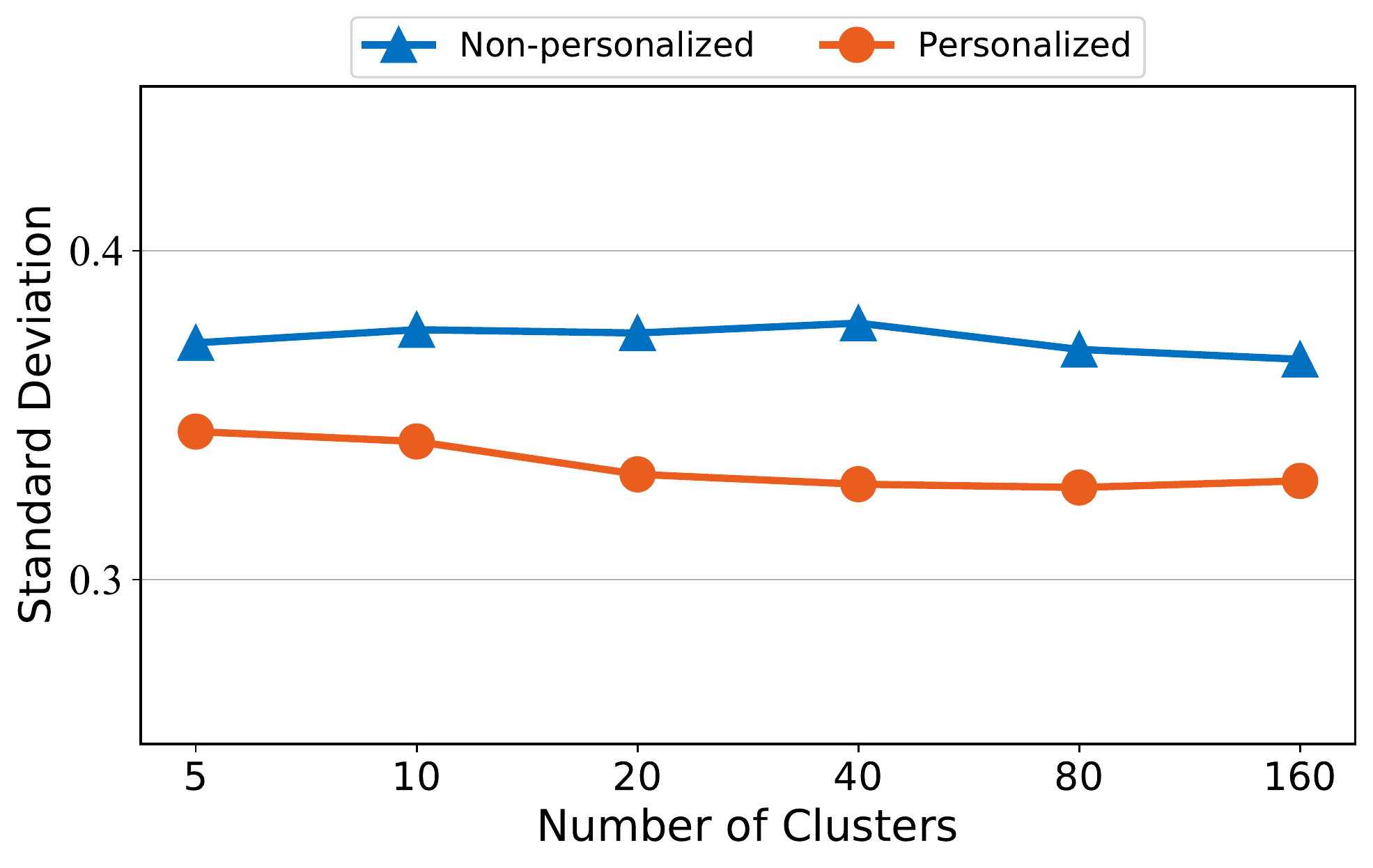}}
			\caption{The impact of personalization components on fairness under different cluster size. The only difference between personalized and non-personalized situations is the former trains local models while the latter does not. As the number of clusters increases, the degree of personalization deepens, and the benefits to fairness increase.}
			\label{fig:fairness}
		\end{figure}
		
		Further ablation study indicates more clearly the importance of personalized local models on the fairness of recommendation results, as shown in Figure \ref{fig:fairness}. The fairness metric of a global model trained with ClusSamp and FastAgg but no ClusUp is depicted in blue, whereas the fairness of local models trained using the extra personalization method is depicted in orange. The obvious gap between them indicates that personalization can effectively improve the fairness of recommendations. The obvious disparity between them suggests that personalization can significantly improve recommendation fairness. Furthermore, the granularity of personalization (as denoted by the number of clusters) also has a considerable impact on fairness: as personalization increases, recommendations generally become more balanced.
		\begin{figure}
			\centering
			\includegraphics[width=8cm]{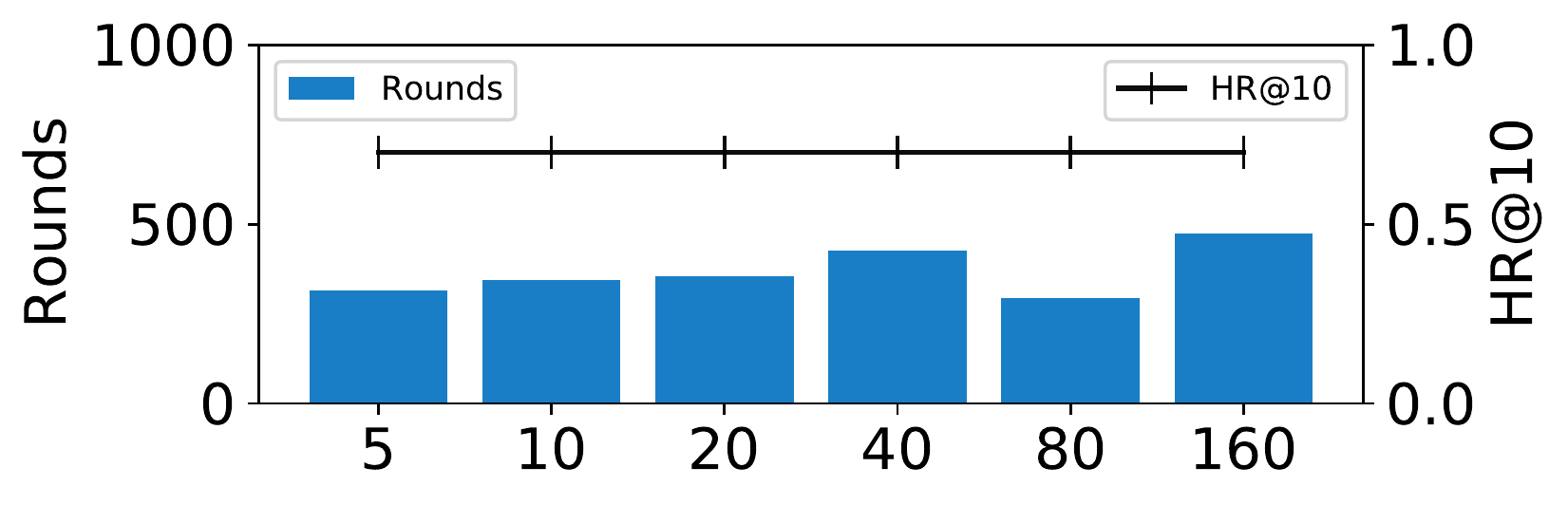}\vspace{1pt} 
			\includegraphics[width=8cm]{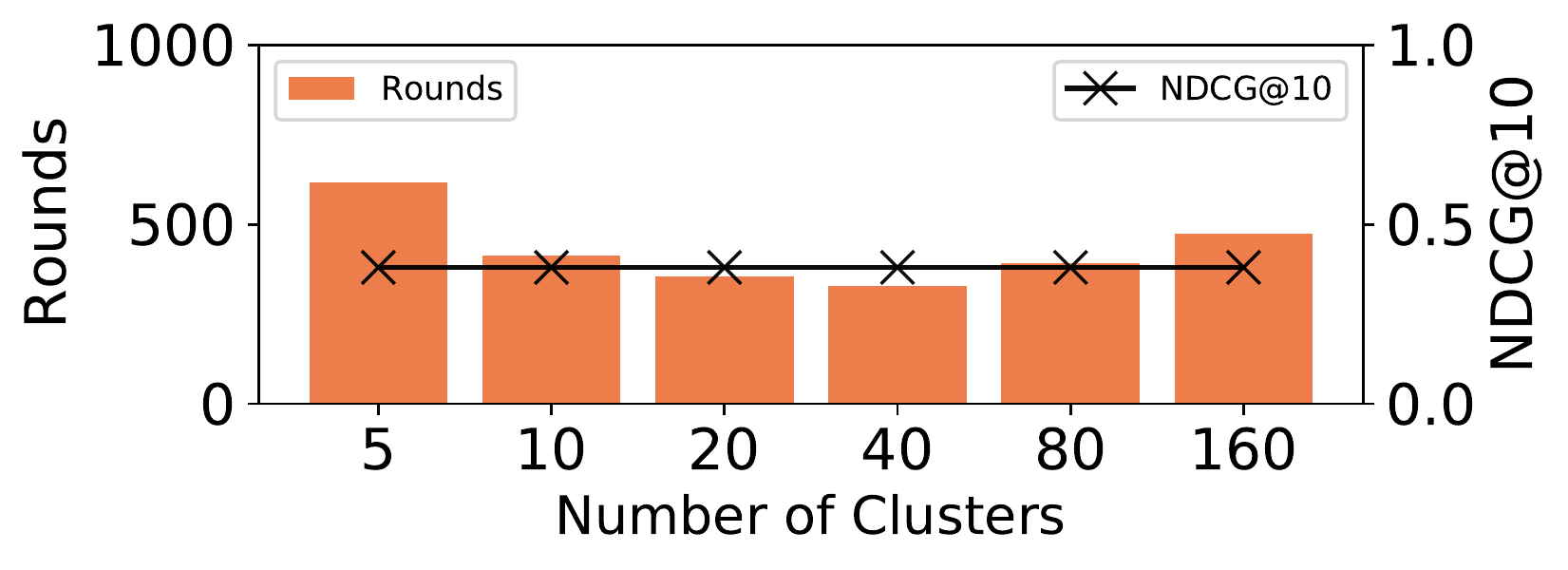}\vspace{1pt}
			\\
					\vspace{-0.3cm}
			\caption{Rounds required for best recommendation performance of Cali3F-GMF on ML-100K dataset. Different number of clusters has great influence on the number of rounds to achieve convergence but little influence on the recommendation effect.}
			\label{fig:rounds}
		\end{figure}
		On the other hand, we can see from Figure 5 that, within a certain range, the number of clusters has little influence on the final recommendation effect. It does, however, have an obvious impact on the number of convergent rounds required to reach the optimal recommendation effect. There is no discernible pattern but the optimal number of rounds in the results is compatible with the unused metadata of ML-100K which can be at best partitioned into about 30 clusters on the offline evaluation. Other datasets' results are not shown because of space limits, but they support this observation as well. This result further proves that our framework can achieve optimal performance without sacrificing privacy.  
		
		\FloatBarrier
		\section{Conclusion}
		We proposed Cali3F, a framework for improving recommendation performance fairness across federated clients while maintaining a fast convergence speed. Cali3F integrates neural networks into the federated collaborative filtering recommendation framework and adopts MTL to train local and global models simultaneously through an adaptive $l_2$ norm regularization term. To address the problem of doubling computing demand, we propose a novel delegates sampling strategy exploiting user profile representativeness, as well as a novel block update component based on user profile similarity, to accelerate the distributed training, which achieves quite a considerable recommendation performance even at the very early stages. In terms of convergence speed and fairness, our experimental results on three real-world datasets show that our framework outperforms federated learning baselines. The theoretical analysis of our proposed framework's ability will be explored in future work. We will also work on how it may be used to improve the robustness.
		\section*{Acknowledgement}
		This paper is supported by the Key Research and Development Program of Guangdong Province under grant No.2021B0101400003. Corresponding author is Jianzong Wang from Ping An Technology (Shenzhen) Co., Ltd (jzwang@188.com).

		\balance
		\footnotesize
		\bibliography{refs}
		\bibliographystyle{IEEEbib}
		
	\end{document}